\documentclass[reprint,prl,twocolumn,superscriptaddress,showpacs,showkeys,floatfix]{revtex4}
\pdfoutput=1

\usepackage{graphicx}

\newcommand{\ba}{\begin{eqnarray*}}
\newcommand{\ea}{\end{eqnarray*}}
\newcommand{\be}{\begin{equation}}
\newcommand{\ee}{\end{equation}}
\newcommand{\bd}{\begin{displaymath}}
\newcommand{\ed}{\end{displaymath}}
\newcommand{\Eq}[1]{Eq.\,#1}
\newcommand{\Fig}[1]{Fig.\,#1}

\newcommand{\amu}{a_\mu}
\newcommand{\alhvp}{a_{l}^\mathrm{hvp}}
\newcommand{\aehvp}{a_{e}^\mathrm{hvp}}
\newcommand{\amuhvp}{a_{\mu}^\mathrm{hvp}}
\newcommand{\atauhvp}{a_{\tau}^\mathrm{hvp}}
\newcommand{\albarhvp}{a_{\overline{l}}^\mathrm{hvp}}
\newcommand{\amubarhvp}{a_{\overline{\mu}}^\mathrm{hvp}}
\newcommand{\hvp}{hvp}
\newcommand{\plotsize}{0.40\textwidth}
\newcommand{\plotgap}{0.04\textwidth}
\newcommand{\plotangle}{0}

\begin{document}

\title{Two-flavor QCD correction to lepton magnetic moments\\at leading-order in the electromagnetic coupling}

\date{June 8, 2011}

\author{Xu Feng}
\altaffiliation{Current address:\ KEK Theory Center, High Energy Accelerator Research Organization (KEK), Tsukuba 305-0801, Japan}
\affiliation{NIC, DESY, Platanenallee 6, D-15738 Zeuthen, Germany}
\affiliation{Universit\"at M\"unster, Institut f\"ur Theoretische Physik, Wilhelm-Klemm-Strasse 9, D-48149, Germany}

\author{Karl Jansen}
\affiliation{NIC, DESY, Platanenallee 6, D-15738 Zeuthen, Germany}

\author{Marcus Petschlies}
\affiliation{Institut f\"ur Physik, Humboldt-Universit\"at zu Berlin, D-12489, Berlin, Germany}

\author{Dru B.\ Renner}
\altaffiliation{Current address:\ Jefferson Lab, 12000 Jefferson Avenue, Newport News, VA 23606, USA}
\affiliation{NIC, DESY, Platanenallee 6, D-15738 Zeuthen, Germany}

\collaboration{ETMC Collaboration}
\noaffiliation

\begin{abstract}
We present a reliable nonperturbative calculation of the QCD
correction, at leading order in the electromagnetic coupling, to the
anomalous magnetic moment of the electron, muon, and tau leptons using
two-flavor lattice QCD.  We use multiple lattice spacings, multiple
volumes, and a broad range of quark masses to control the continuum,
infinite-volume, and chiral limits.  We examine the impact of the
commonly ignored disconnected diagrams and introduce a modification to
the previously used method that results in a well-controlled lattice
calculation.  We obtain $1.513\,(43)\cdot 10^{-12}$, $5.72\,(16)\cdot
10^{-8}$, and $2.650\,(54)\cdot 10^{-6}$ for the leading-order
two-flavor QCD correction to the anomalous magnetic moment of the
electron, muon, and tau, respectively, each accurate to better than
$3\%$.
\end{abstract}



\maketitle

\section{Introduction}

The experimental~\cite{Bennett:2006fi} and
theoretical~\cite{Jegerlehner:2009ry} determinations of the anomalous
magnetic moment of the muon $a_\mu$ have both reached an accuracy that
is better than six parts per million.  This high precision reveals a
discrepancy of over 3 standard deviations ($3\sigma$), which
raises the possibility of physics beyond the standard model.  However,
the dominant error in the theory computation is due to hadronic
effects that are currently not calculated but are instead either
separately measured or simply modeled.  This obscures the significance
of the $3\sigma$ effect and makes it difficult to improve the accuracy
of the Standard Model calculation.

In this Letter, we present a reliable lattice QCD determination of the
leading-order hadronic correction for the muon, $\amuhvp$, which is
the single largest source of error in the theory calculation of
$a_\mu$.  Additionally, we calculate the leading-order corrections
$\aehvp$ for the electron and $\atauhvp$ for the tau, achieving an
accuracy of better than $3\%$ for each.  This was accomplished by
introducing a modification of the existing method that results in a
significantly more well-controlled calculation.  After examining all
sources of systematic error and performing our own extraction of the
two-flavor contribution to the experimental measurements, we find
agreement for all three charged leptons in the standard model.

Our current computation is performed in two-flavor QCD, but the
technique presented in this work is readily generalized to a realistic
four-flavor calculation that is already under way~\cite{Baron:2010bv}.
The precision of our calculation and the prospects for improving it
demonstrate that lattice QCD can realistically provide a
first-principles determination of the leading-order hadronic
contributions to the magnetic moments of the standard model leptons.

\section{Leading-order Hadronic Correction}

The anomalous magnetic moment $a_l$ of a lepton $l$ can be written as
a perturbative expansion in the electromagnetic coupling $\alpha$.
Contributions from QCD first occur at the order $\alpha^2$ and can be
written as~\cite{Blum:2002ii}
\be
\label{alhvp}
\alhvp = \alpha^2\!\! \int_0^{\infty}\!\!\!\! dQ^2 \frac{1}{Q^2} w(Q^2/m_l^2)\, \Pi_R(Q^2)\,,
\ee
where $m_l$ is the mass of the lepton, $Q$ is the Euclidean momentum
and $w(Q^2/m_l^2)$ is a known function.  The combination $\Pi_R(Q^2) =
\Pi(Q^2) - \Pi(0)$ is the renormalized hadronic vacuum polarization
function $\Pi(Q^2)$, which is defined shortly.  The weight function
$w(Q^2/m_l^2)$ vanishes as $(Q^2)^{-2}$ for large $Q^2$.  This ensures
that the integral above is dominated by the low $Q^2$ region, making
it clear that $\alhvp$ must be evaluated nonperturbatively.

\section{Experimental Determination}

The electron and muon magnetic moments have been measured in dedicated
experiments~\cite{Hanneke:2008tm,Bennett:2006fi}.\ To compare to the
standard model prediction, the leading-order hadronic correction is
determined by using unitarity and causality to relate the expression
in \Eq{\ref{alhvp}} to
\be
\label{alhvpex}
\alhvp = \alpha^2\!\! \int_0^{\infty}\!\!\!\! ds\, \frac{1}{s} w^\prime(s/m_l^2)\, R(s)\,.
\ee
Here $w^\prime$ is another known weight function and $R(s)$ is the
ratio of the hadronic cross section $\sigma(e^+
e^-\!\rightarrow\!\mathrm{hadrons})$ to the leptonic cross section
$\sigma(e^+ e^-\!\rightarrow\!  \mu^+\mu^-)$.  The determination of
$R(s)$ relies on the results of many experiments, and the integral in
\Eq{\ref{alhvpex}} has been evaluated by several groups, most
recently~\cite{Jegerlehner:2009ry, Davier:2010nc, Jegerlehner:2011ti,
  Hagiwara:2011af}.  Additionally, there are higher-order corrections,
including the so-called light-by-light contribution, which is
difficult to measure and is modeled instead.

Our calculation is performed in QCD with only up and down quarks, so
we need to extract the two-flavor contribution to $\alhvp$.
Inevitably, this introduces some ambiguity.  For the purposes of
comparing to our current two-flavor calculation, we adopt the simple
procedure of rescaling the contribution to the integral in
\Eq{\ref{alhvpex}} from the energy regions between quark thresholds by
the value of $\sum_f Q_f^2$, where the sum runs over only the active
quark flavors for that region and the electric charges of the quarks
are $eQ_f$.  This neglects the very small changes due to the running
of the QCD coupling, it ignores small off-diagonal contributions
proportional to $Q_f Q_{f^\prime}$, and it disregards any
complications at the flavor thresholds.  These are all caveats that we
must accept in the current comparisons but that will be eliminated in
our ongoing four-flavor computation.

Using the results from \cite{Jegerlehner:2008zz, Jegerlehner:1996ab},
we extract the two-flavor contributions to $\alhvp$ along the lines
just described, giving
$a_{e,N_f=2}^\mathrm{\hvp,ex} = 1.547\,(36)\cdot 10^{-12}$, 
$a_{\mu,N_f=2}^\mathrm{\hvp,ex} = 5.660\,(47)\cdot 10^{-8}$, and
$a_{\tau,N_f=2}^\mathrm{\hvp,ex} = 2.638\,(88)\cdot 10^{-6}$.
The errors result from propagating just those of
\cite{Jegerlehner:2008zz, Jegerlehner:1996ab}.  The systematic error
due to extracting the two-flavor contribution is likely larger than
these uncertainties and must be taken into consideration when
comparing our calculation to these estimates.
 
\section{Lattice QCD Calculation}

The leading-order hadronic correction, $\alhvp$, is the order
$\alpha^2$ contribution in a perturbative QED expansion of $a_l$ but
it must be treated nonperturbatively in QCD.  To this order in the QED
coupling, the QCD corrections only modify the photon propagator.
These contributions can be formally summed to all orders giving the
hadronic vacuum polarization tensor,
\bd
\Pi_{\mu\nu}(Q) = \!\! \int\!\! d^4\!X\, e^{iQ\cdot X} \langle \Omega | T J_\mu(X) J_\nu(0) | \Omega \rangle\,.
\ed
The current $J_\mu = \sum_f Q_f \overline{q}_f \gamma_\mu q_f$ is the
hadronic component of the electromagnetic current and the sum runs
over all relevant quark flavors.  The current $J_\mu$ is conserved,
consequently this correlation function satisfies a Ward identity that
allows us to write $\Pi_{\mu\nu}$ in terms of a single scalar function
of $Q^2$ as
\bd
\label{vacpolfun}
\Pi_{\mu\nu}(Q) = (Q_\mu Q_\nu - Q^2 \delta_{\mu\nu} ) \Pi(Q^2)\,.
\ed
Note that both $\Pi_{\mu\nu}$ and $\Pi$ are calculated directly in
Euclidean space without any analytic continuation.

We use standard lattice QCD techniques to calculate the
vacuum-to-vacuum matrix element $\langle \Omega | T J_\mu(x) J_\mu(y)
| \Omega \rangle$.  The functional integral that is implicit in this
correlation function is evaluated stochastically using the results of
the European Twisted Mass Collaboration~\cite{Baron:2009wt}.  We have
used two lattice spacings, $a=0.079$ and $a=0.063~\mathrm{fm}$, to
examine lattice cutoff corrections.  Two finite-volume studies were
performed to check for finite-size effects.  The up and down quark
masses $m_q$ are equal and are parametrized in terms of the
pseudoscalar meson mass $m_{PS}$, with $m_q \propto m_{PS}^2$ in the
chiral limit.  As is common, we use heavier-than-physical quark masses
and then take the limit as $m_{PS}$ approaches the physical pion mass
$m_\pi$.  This was done by studying the dependence on $m_{PS}$ over
the range from $650$ to $290~\mathrm{MeV}$.  The so-called
disconnected diagrams, ignored in all previous calculations, were
included for almost half of the ensembles used in this work and are
accounted for as a systematic error along with those from the
continuum, infinite-volume and physical quark-mass limits.  The
additional details are standard and deferred to a later publication.

Apart from variations in how the lattice calculation of $\Pi(Q^2)$ is
matched to a smooth function, the method used so far in all
calculations~\cite{Blum:2002ii, Gockeler:2003cw, Aubin:2006xv,
  Brandt:2010ed} proceeds by numerically integrating \Eq{\ref{alhvp}}
directly to form $\alhvp$.  In our calculation, we parametrize
$\Pi(Q^2)$ over the entire range of $Q^2$ that is determined from our
lattice computation.  The presence of the lattice cutoff and the
restriction to finite volume induce an ultraviolet cutoff
$Q^2_\mathrm{uv}$ proportional to $1/a^2$ and an infrared cutoff
proportional to $1/L^2$.  Extrapolating the functional form for
$\Pi(Q^2)$ to $Q^2=0$~\cite{Renner:2011ff}, we numerically evaluate the
integral from $Q^2=0$ to $Q^2=Q_\mathrm{uv}^2$.  This is done without
any use of perturbation theory, giving a completely nonperturbative
evaluation of $\alhvp$.  The systematic error caused by extrapolating
to $Q^2=0$ is eliminated as $L$ is taken large and the error due to
truncating the integral at $Q_\mathrm{uv}^2$ is removed as $a$ goes to
zero.  Thus both effects are automatically accounted for as part of
the corresponding systematic errors.

\begin{figure}
\includegraphics[width=\plotsize,angle=\plotangle]{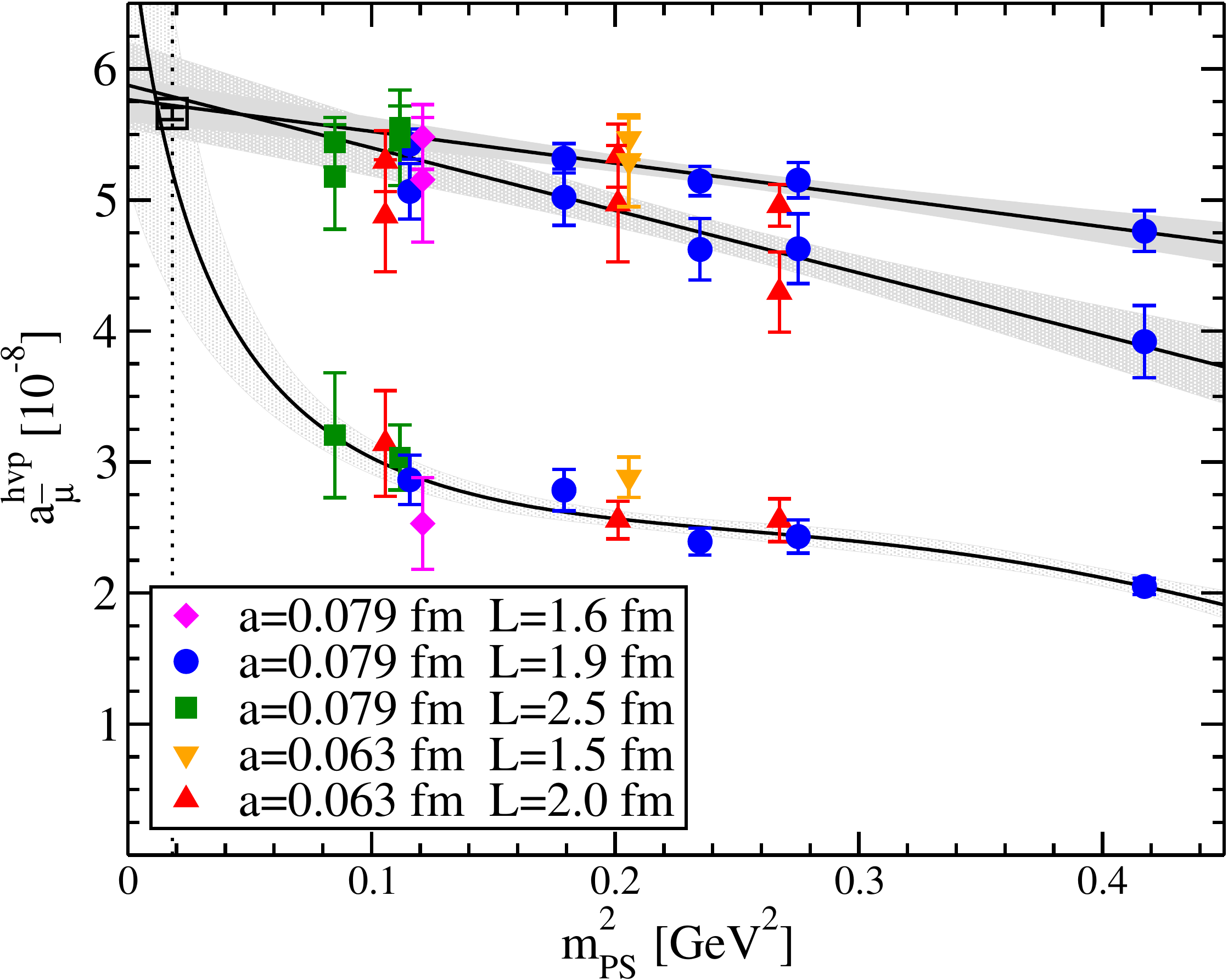}\hspace{\plotgap}
\caption{Comparison of methods for $\amuhvp$.  The upper set of points
  are the results for $\amubarhvp$ using $H=m_V$, the middle set use
  $H=f_V$, and the lower set correspond to the standard method,
  formally $H=1$.  The two lines are linear extrapolations of
  $\amubarhvp$ and the curve is the phenomenological extrapolation of
  $\amuhvp$.  The three methods agree at the physical point, denoted
  by the dashed line, and agree with the estimated two-flavor
  contribution to the experimental value.}
\label{amucomp}
\end{figure}
Our results for the muon using this method, which we deem the
standard method, are shown as the lowest set of points in
\Fig{\ref{amucomp}}.  Consistent with all other lattice calculations
of $\amuhvp$, we find that the values calculated at $m_{PS}$ heavier
than $m_\pi$ are significantly lower than the experimentally measured
value and apparently rise rapidly only when $m_{PS}$ approaches quite
near the physical value $m_\pi$.

We attribute this behavior to the contributions of the lowest-lying
vector mesons.  The rho, omega, and phi mesons account for over $80\%$
of the fully measured $\amuhvp$~\cite{Jegerlehner:2008zz}.  Any
description of the vector-meson contribution to $\alhvp$ will depend
on the mass $m_V$ and a variety of dimensionless couplings.  Without
loss of generality, we focus on just those models based on $m_V$ and
the electromagnetic coupling $g_V$ with $\langle \Omega | J_\mu | V,
\epsilon \rangle = m_V^2 g_V \epsilon_\mu / \sqrt{2}$.  The coupling
shows up quadratically and dimensional analysis then results in a
vector-meson contribution of $a_{l,V}^\mathrm{hvp} = g_V^2 f(
m_l^2/m_V^2 )$.  Additionally, $f(m_l^2/m_V^2)$ should vanish for
$m_l\rightarrow 0$ and $m_V\rightarrow\infty$.  Thus on rather general
grounds we expect $a_{l,V}^\mathrm{hvp} \approx C g_V^2 m_l^2 / m_V^2$
with a model-dependent constant $C$.

\begin{figure}
\includegraphics[width=\plotsize,angle=\plotangle]{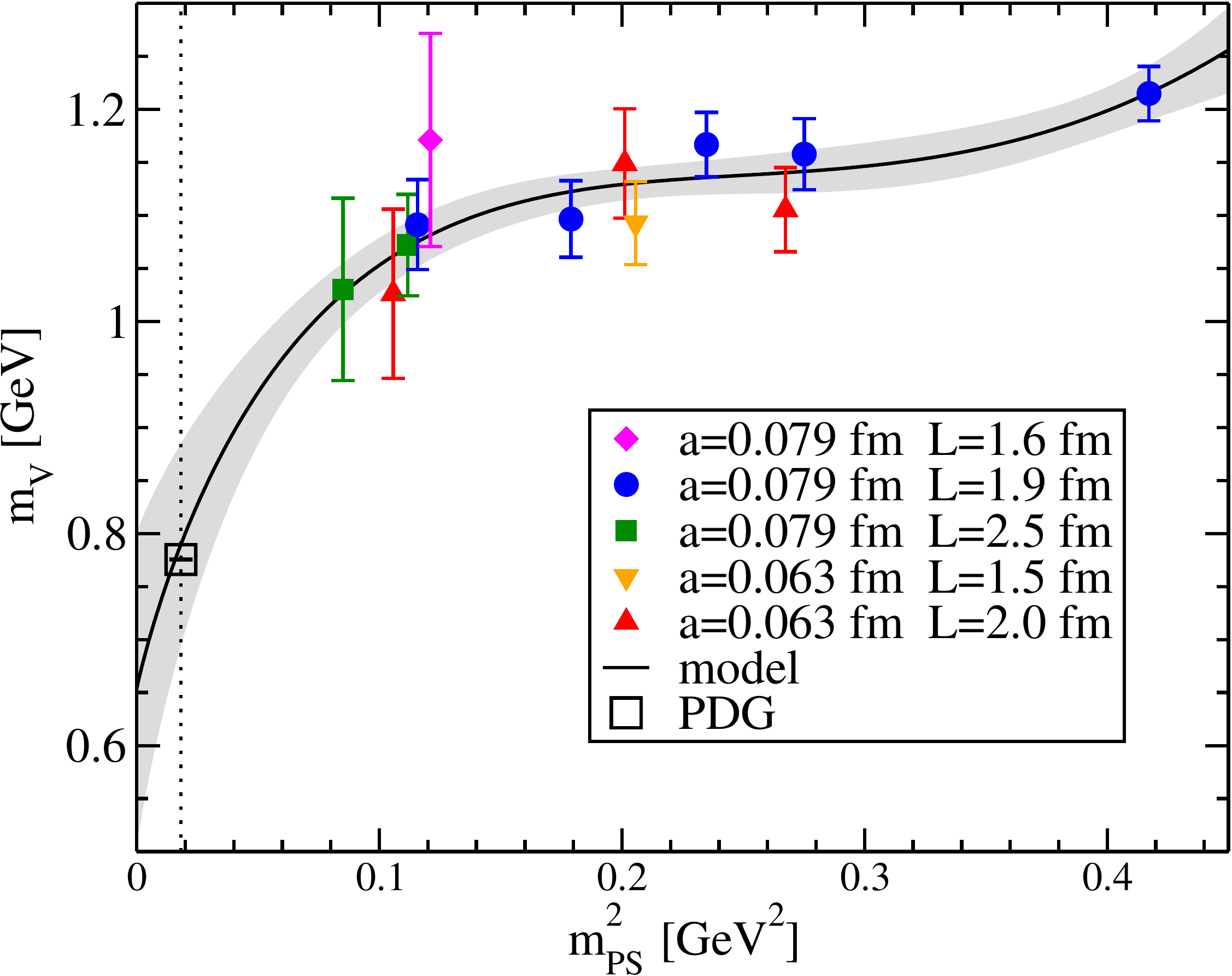}\hspace{\plotgap}
\caption{Phenomenological model for $m_V$.  A model function is used
  to parametrize both our lattice calculation of $m_V$ and the PDG
  value of the physical $m_\rho$.  This model is only used to
  illustrate the difficulties in the standard method.}
\label{mv}
\end{figure}
These expectations can be combined with our lattice calculation of
$m_V$ and $g_V$.  As shown in \Fig{\ref{mv}}, we find that $m_V$
decreases moderately with decreasing $m_{PS}$ but the values from our
calculation are still rather high compared to the experimental result
$m_\rho$.  Thus at some point a rapid decrease in $m_V$ must occur.
In contrast, $g_V$, not shown but well fit by $g_V=0.29(1)-0.09(2)\,
m_{PS}^2$, has a mild dependence on $m_{PS}$ and extrapolates smoothly
to the experimental value $g_\rho$.  When combined with the model
expectation $a_{\mu,V}^\mathrm{hvp} \propto g_V^2 / m_V^2$, the
behavior of $\amuhvp$ in \Fig{\ref{amucomp}} becomes plausible.  The
values of $\amuhvp$ are lower than the experimental value and vary
moderately for the region of $m_{PS}$ covered in our calculation.
Only at lighter values of $m_{PS}$ do we expect a sharp increase in
$\amuhvp$.

We can make these observations more precise, at the expense of
introducing model dependence, by considering the tree-level form for
the vector-meson contribution $a_{l,V}^\mathrm{hvp}$ as given from
effective field theory~\cite{Aubin:2006xv}.  This gives a specific
result for $f(m_l^2/m_V^2)$ that we combine with our calculation of
$m_V$ and $g_V$ to construct a model-dependent extrapolation of the
results for $\amuhvp$.  Additionally, constraining $m_V$ to approach
$m_\rho$ as shown in \Fig{\ref{mv}} gives the lowest-lying curve in
\Fig{\ref{amucomp}}.  The apparent agreement with the physical value
for $\amuhvp$ increases the plausibility that our explanation is
correct.  However, this construction does not provide a reliable means
of extrapolating our results to the physical $m_\pi$ but instead
serves to illustrate the apparently strong $m_{PS}$ dependence in the
standard method.

The difficulties encountered in the standard method can be traced to
the occurrence of two distinct scales, $m_l$ and $m_V$.  Apart from
any model, this is relevant because $\alhvp$ is made dimensionless at
the expense of introducing an external scale $m_l$ that is completely
unrelated to the scales of QCD.  Based on this observation, we define
the following class of observables:
\be
\label{albarhvp}
\albarhvp = \alpha^2\!\! \int_0^{\infty}\!\!\!\! dQ^2 \frac{1}{Q^2} w((Q^2/m_l^2) (H^2_\mathrm{phys} / H^2))\, \Pi_R(Q^2)
\ee
where $H$ is any hadronic quantity, understood to be a function of
$m_{PS}$, and $H_\mathrm{phys}$ is its physical value.  The natural
choice for our calculation is $H=m_V$, but any choice produces a new
modified quantity that has the same physical limit as $\alhvp$.  This
follows simply by construction because $H(m_{PS}\rightarrow m_\pi) =
H_\mathrm{phys}$.  The standard method can be formally reproduced by
the choice $H=1$, but choosing a dimensionful scale has the additional
advantage that the explicit dependence on the lattice spacing is
eliminated.  At the same time, the renormalization condition that
defines the physical limit is now given by the dimensionless ratio
$m_l/H_\mathrm{phys}$ rather than $m_l$ alone.

The calculation of $\amubarhvp$ using $H=m_V$ and $H=f_V$, the
vector-meson decay constant given by $f_V = m_V g_V$, are shown in
\Fig{\ref{amucomp}}.  All three extrapolations agree with each other
and with the estimated two-flavor contribution to the experimental
measurement of $\amuhvp$.  The results for the new method show a
significantly milder dependence on $m_{PS}$.  This can be understood
using the model considerations earlier.  Specifically for $H=m_V$, we
expect a vector-meson contribution of $a_{{\overline l},V} \approx C
g_V^2 m_l^2/m_\rho^2$, in which only the mild $m_{PS}$ dependence of
$g_V$ now enters.  The demonstration that $H=f_V$ results in similar
improvements illustrates that any quantity sensitive to $m_V$ will
likely yield a well-controlled observable $\albarhvp$.

Without regard to any particular model or the experimental
measurements, we can examine the relative merits of the standard and
modified methods. Using the muon as an example, the shift between
linear and quadratic extrapolations for $\amubarhvp$ (using $H=m_V$)
is $1.7\%$, which is only a $0.6\,\sigma$ effect.  The same results
for $\amuhvp$ are $17\%$ and $3.5\,\sigma$, indicating the presence of
noticeably more curvature in the standard approach.  In this case,
cubic fits are required and give an extrapolated value of
$4.1\,(1.5)\cdot 10^{-8}$, which agrees with the more precise value of
$5.72\,(16)\cdot 10^{-8}$ that results from extrapolating
$\amubarhvp$.  The same pattern holds for the electron and tau; thus,
the lattice calculation itself provides direct evidence that our
modified method has a smoother approach to the physical limit leading
to a more accurate calculation.

\begin{figure}
\includegraphics[width=\plotsize,angle=\plotangle]{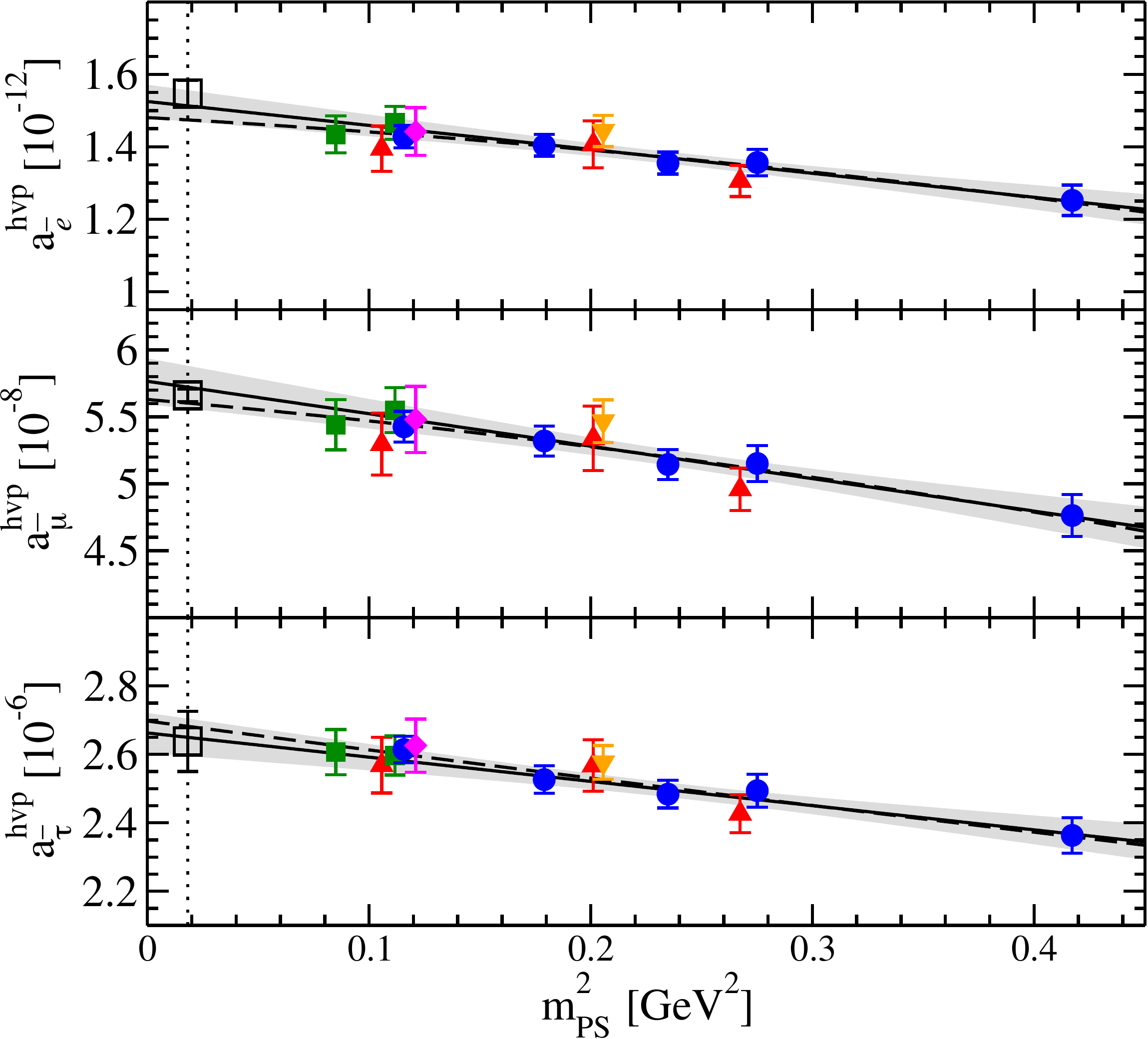}\hspace{\plotgap}
\caption{Calculation of $\albarhvp$ for all three $l=e$, $\mu$ and
  $\tau$.  The meaning of the symbols is the same as in
  \Fig{\ref{amucomp}}.  We show the results from our improved method
  using $H=m_V$.  The results are extrapolated linearly (solid line
  with error band) and quadratically (dashed line) to the physical
  point and agree with the two-flavor contribution extracted from the
  experimental measurements.}
\label{albarplot}
\end{figure}
Taking the modified method with $H=m_V$ as our definition of
$\albarhvp$, we calculate all three $l=e$, $\mu$ and $\tau$.  These
results are shown in \Fig{\ref{albarplot}}, and the extrapolated
values at the physical point are
\ba
a_{e,N_f=2}^\mathrm{hvp} &=& 1.513\,(43)\cdot 10^{-12} \\
a_{\mu,N_f=2}^\mathrm{hvp} &=& 5.72\,(16)\cdot 10^{-8} \\
a_{\tau,N_f=2}^\mathrm{hvp} &=& 2.650\,(54)\cdot 10^{-6}\,.
\ea
The quoted errors are due to the stochastic integration only.  We do
not find any statistically meaningful uncertainties due to lattice
artifacts, finite-size effects, the extrapolation in $m_{PS}$ or the
exclusion of the disconnected diagrams.  At some higher precision
these effects will be relevant, but there is no sign that they are
significant at the few percent level of our current calculation.

\section{Conclusions and Outlook}

We have performed the first lattice QCD calculation of the
leading-order QCD correction to the anomalous magnetic moments
$\alhvp$ that included dynamical quarks, examined lattice artifacts,
checked finite-size effects and studied the disconnected diagrams.  We
examined the pitfalls of the standard method for calculating $\alhvp$
and introduced a modification that creates a dimensionless quantity
$\albarhvp$ composed of hadronic scales only.  This quantity has the
same physical limit as $\alhvp$ but has a mild approach to that limit
that is now well controlled.  This allowed us to calculate the
leading-order correction for all three charged leptons with an
accuracy better than $3\%$, reproducing our estimate of the two-flavor
contributions to the experimental measurements.

The calculation was done using two-flavor QCD, which is the most
significant systematic error.  To resolve this, we are currently
starting a four-flavor calculation.  This will eliminate any ambiguity
regarding the extraction of the two-flavor experimental value.  When
combined with further anticipated improvements, the modified method
presented here should produce a result precise enough to replace the
experimentally estimated $\alhvp$ with a complete first-principles QCD
calculation and eliminate this source of ambiguity in the current
$3\sigma$ discrepancy in $\amu$.

\begin{acknowledgments}
We thank our fellow members of ETMC for their constant collaboration.
We are grateful to the John von Neumann Institute for Computing (NIC),
the J{\"u}lich Supercomputing Center and the DESY Zeuthen Computing
Center for their computing resources and support.  This work has been
supported in part by the DFG Sonderforschungsbereich/Transregio
SFB/TR9-03, the DFG project Mu 757/13, and the U.S. DOE under Contract
No. DE-AC05-06OR23177.
\end{acknowledgments}


\bibliography{letter}
\bibliographystyle{h-physrev}

\end{document}